\documentclass{article}
\usepackage{fullpage}
\usepackage{graphics,psfrag,epsfig}
\usepackage{amsfonts,latexsym,eucal,amsmath,amsthm,amssymb}

\begin{document}

\newcommand{\rum}{\rule{0.5pt}{0pt}}
\newcommand{\rub}{\rule{1pt}{0pt}}
\newcommand{\rim}{\rule{0.3pt}{0pt}}
\newcommand{\numtimes}{\mbox{\raisebox{1.5pt}{${\scriptscriptstyle \times}$}}}
\newcommand{\optprog}[2]
{%
  \noindent\mbox{}\\[0cm]
  \noindent\fbox{%
  \begin{minipage}{0.955\linewidth}
    \mbox{}\\[-0.5cm]
    #1\\[#2]
  \end{minipage}
  }
  \noindent\mbox{}\\[-0.2cm]
}

\renewcommand{\refname}{References}

\twocolumn[%
\begin{center}
{\Large\bf A mathematical definition of ``simplify" \rule{0pt}{13pt}}\par
\bigskip
Craig Alan Feinstein \\ {\small\it 2712 Willow Glen Drive, Baltimore, Maryland
21209\rule{0pt}{13pt}}\\ \raisebox{-1pt}{\footnotesize E-mail: cafeinst@msn.com,
BS"D}\par
\bigskip\smallskip
{\small\parbox{11cm}{%
\bigskip \noindent \textbf{Abstract:} Even though every mathematician knows intuitively what it means to ``simplify" 
a mathematical expression, there is still no universally accepted rigorous mathematical 
definition of ``simplify". In this paper, we shall give a simple and plausible definition 
of ``simplify" in terms of the computational complexity of integer functions. We shall also use this definition
to show that there is no deterministic and exact algorithm which can compute the permanent 
of an $n \times n$ matrix in $o(2^n)$ time.

\bigskip \noindent \textbf{Disclaimer:} This article was authored
by Craig Alan Feinstein in his private capacity. No official support or endorsement by
the U.S. Government is intended or should be inferred.\rule[0pt]{0pt}{0pt}}}
\bigskip
\end{center}]{%

\section{Introduction}

In 2013, the author asked the following quesiton titled ``Is there a `mathematical' definition of `simplify'?"
on the popular mathematics website MathOverflow.net \cite{MO13}:

``Every mathematician knows what `simplify' means, at least intuitively. Otherwise, he or 
she wouldn't have made it through high school algebra, where one learns to `simplify' expressions like    
$x(y+x)+x^2(y+1+x)+3(x+3)$. But is there an accepted rigorous `mathematical' definition of `simplify' not 
just for algebraic expressions but for general expressions, which could involve anything, 
like transcendental functions or recursive functions? If not, then why? I would think 
that computer algebra uses this idea."

The answers there indicated that even though every mathematician knows intuitively what ``simplify"
means, there is still no universally accepted definition of ``simplify". In fact, one of the answers (by Henry Cohn) indicated
that ``In full generality, there provably isn't any method for complete simplification". (He was referring to elementary functions of a real variable.)
In this paper, we shall give a simple and plausible definition of ``simplify" in terms of the computational complexity of integer functions.
We shall also use this definition to show that there is no deterministic and exact algorithm which can compute the permanent 
of an $n \times n$ matrix in $o(2^n)$ time.

\section{A definition of ``simplify"}

Consider the following definition of ``simplify":

\bigskip\noindent\textbf{Definition:} An algebraic expression (recursive or non-recursive) for a function $f: \mathbb Z \rightarrow \mathbb Z$ 
\textit{cannot be simplified} if there is no other algebraic expression for $f$ which can be computed faster.

\bigskip\noindent For example, the expression $xw+yz+xz+yw$ can be simplified to $(x+y)(w+z)$, since
computing $(x+y)(w+z)$ takes only one multiplication and two additions, while computing $xw+yz+xz+yw$ 
takes four multiplications and three additions. And we can also see clearly that the expression $(x+y)(w+z)$ cannot 
be simplified.

As another example, let $f: \mathbb Z \rightarrow \mathbb Z$ be the function which satisfies the recursive formula,  
$f(n)=f(n-1)+1$ and $f(0)=0$. This recursive formula can be simplified to $f(n)=n$, since computing the recursive formula for $f$
takes $\Theta(n)$ time, while computing the formula $f(n)=n$ is trivial. And the formula $f(n)=n$ clearly cannot be simplified. 

And let $f: \mathbb N \rightarrow \mathbb N$ be the function which satisfies the recursive formula, 
$f(n)=f(n-1)+f(n-2)$ and $f(1)=f(2)=1$, the Fibonacci sequence. This recursive formula can be simplified, since it is possible to 
prove that $f(n)$ equals $\phi^n/\sqrt{5}$ rounded to the nearest integer,
where $\phi=(1+\sqrt{5})/2$, which can be computed exponentially faster than the recursive formula can be computed \cite{b:Fib}.

\section{Computing the permanent of a matrix}

Let $A=(a_{ij})$ be a matrix of integers. The permanent of $A$ is defined as:
$$\mbox{perm}(A)=\sum_{\sigma \in S_n}\prod_{i=1}^n a_{i\sigma(i)},$$
where $S_n$ is the symmetric group \cite{b:Perm}. The fastest known deterministic and exact algorithm 
which computes the permanent of a matrix was first published in 1963 and has a running-time of 
$\Theta^*(2^n)$ \cite{b:Rys}. It is still considered an open problem by the mathematics and computer science community whether this 
time can be beaten. Now consider the following theorem and proof, which we shall discuss afterwards:

\bigskip\noindent\textbf{Theorem:} There is no deterministic and exact algorithm which can compute the permanent of 
an $n \times n$ matrix in $o(2^n)$ time.

\bigskip\noindent\textit{Proof:} For any row $i$, the permanent of matrix $A$ satisfies the recursive formula 
$$\mbox{perm}(A)=\sum_{j=1}^n a_{ij}\cdot\mbox{perm}(A_{ij}^\#)$$
and $\mbox{perm}([a_{11}])=a_{11}$, where $A_{ij}^\#$ is the $(n-1) \times (n-1)$ matrix that 
results from removing the $i$-th row and the $j$-th column from $A$. This formula cannot be 
simplified, so the fastest algorithm for computing the permanent of a matrix is to apply this 
recursive formula to matrix $A$. Since this involves recursively evaluating the permanent of 
$\Theta(2^n)$ submatrices of $A$, each corresponding to a subset of the $n$ columns of $A$, we 
obtain a lower bound of $\Theta(2^n)$ for the worst-case running-time of any deterministic and 
exact algorithm that computes the permanent of a matrix.\qed

\bigskip\noindent At first, this proof makes sense intuitively, but if one thinks about it a little more, one might become 
skeptical, since one could argue the same for the determinant
of a matrix, that there is no deterministic and exact algorithm which can compute the determinant of an $n \times n$ matrix in $o(2^n)$ time 
(which is known to be false) - for any row $i$, the determinant satisfies the recursive formula
$$\mbox{det}(A)=\sum_{j=1}^n (-1)^{i+j}a_{ij}\cdot\mbox{det}(A_{ij}^\#)$$
and $\mbox{det}([a_{11}])=a_{11}$, which is almost the same as the recursive formula for the permanent of a matrix. 

However, there is a big difference between the two recursive formulas: There are negative signs in the formula for the determinant,
so it is not inconceivable that one might be able to cancel most of its terms out, if one is clever.
And in fact this is the reason why it is possible to compute the determinant of a matrix in polynomial-time:
If one performs elementary row operations on matrix $A$ with pivot $a_{11} \neq 0$, converting it to a matrix $B$
with zeroes in the last $n-1$ entries of column 1, then 
the determinant of $A$ will equal the determinant of $B$ and we will also obtain a simpler formula for the determinant: 
$$\mbox{det}(A)=a_{11}\cdot\mbox{det}(B_{11}^\#).$$
This trick ultimately leads to a polynomial-time algorithm for computing the determinant of a matrix, if one applies it recursively to 
the matrix $B_{11}^\#$, exchanging rows when necessary. 

However, in the case of the permanent of a matrix, no trick like this is possible, since there are only positive signs in its formula. 
To gain some insight as to why this is so, consider the following analogy: Suppose we want to
subtract two large positive numbers with a tiny difference, say $a=12,345,678,907$ and $b=12,345,678,903$. One could compute $a$ minus $b$ by applying the normal
subtraction procedure that one learns in elementary school to each digit of these two numbers, but one does not have to do this; if we let $c=12,345,678,900$,
then we will obtain the same answer by computing $(a-c)$ minus $(b-c)$, which amounts to subtracting only the last digits
of each number, $7$ minus $3$. But there are no short-cuts like this for adding $a$ and $b$, since none of their digits can be cancelled out. And for 
this same reason, it is possible to cancel out lots of terms in the formula for the determinant but not in the formula for the permanent, as 
the elementary row operations which are performed on matrix $A$ when computing its determinant via the algorithm described above are analogous to 
subtracting $c$ from both $a$ and $b$.

But then one might ask, ``The proof above said `This formula cannot be simplified'. But how can I be sure of this?"
The answer to this question is that we know that the above recursive formula for the permanent cannot be simplified, because we have tried every possible
way to simplify it and saw that each way fails: To be specific, we tried to multiply the factors, $a_{ij}$ and $\mbox{perm}(A_{ij}^\#)$, of the summands
together, but we failed since the two factors are completely independent from one another. And we tried adding the summands together,
but we also failed since the factors $a_{ij}$ found in each summand are completely independent from one another and are also completely independent from each 
$\mbox{perm}(A_{ij}^\#)$; furthermore, we found that since $\mbox{perm}(A_{ij}^\#)$ is different in each term, it is impossible to use the distributive 
law to decrease the computational complexity of the recursive expression. And finally, we noticed that the row choice of $i$ is irrelevant in the recursive 
formula for the permanent, so no choice of $i$ is better than any other choice. What other things are there to try that could possibly make the expression simpler? Nothing,
since we have already considered every mathematical operation in the recursive formula for the permanent. Therefore, the 
recursive formula for the permanent cannot be simplified, i.e., it has the best computational complexity of any algebraic expression for the permanent of a matrix. 

This type of reasoning is not new or foreign; it is essentially the same type of reasoning that a 
high school math student uses to simplify algebraic expressions. Also note that only if one is careful in one's analysis and considers every possible way to simplify 
an algebraic expression can one prove that an algebraic expression indeed
cannot be simplified; merely claiming that an algebraic expression cannot be simplified does not make it so. But sometimes it is so obvious that an algebraic expression
cannot be simplified that writing down a full explanation of this is unnecessary. Also, it turns out that one can use similar reasoning to prove 
that there is no deterministic and exact algorithm which solves the Traveling Salesman Problem in polynomial-time \cite{Fei11}.

\section{Conclusion}

While everyone in the mathematics community understands intuitively what ``simplify" in mathematics means, 
there is still no universal definition of ``simplify". In this paper, we have defined ``simplify" in terms of the computational complexity 
of an integer function and have shown that this definition can be used to prove that there is no 
deterministic and exact algorithm which can compute the permanent of an $n \times n$ matrix in $o(2^n)$ time.

}


\smallskip
\begin{thebibliography}{99}\small

\bibitem{MO13} Feinstein, C. A., ``Is there a `mathematical' definition of `simplify'?". https://mathoverflow.net/q/126519/7089

\bibitem{Fei11} Feinstein, C. A., ``The Computational Complexity of the
Traveling Salesman Problem". \textit{Global Journal of Computer
Science and Technology}, Volume 11 Issue 23, December 2011, pp 1-2. https://arxiv.org/abs/cs/0611082

\bibitem{b:Rys} Ryser, H. J. \textit{Combinatorial Mathematics} (Carus Math. Monograph No. 14, 1963).

\bibitem{b:Fib} Chandra, Pravin and Weisstein, Eric W. ``Fibonacci Number." 
From \textit{MathWorld}--A Wolfram Web Resource. http://mathworld.wolfram.com/FibonacciNumber.html

\bibitem{b:Perm} Weisstein, Eric W. ``Permanent." From \textit{MathWorld}--A Wolfram Web Resource. 
http://mathworld.wolfram.com/Permanent.html 

\end{thebibliography}
\end{document}